\documentstyle[12pt,epsf,epsfig]{article}

\title{Cluster Coagulation and Growth Limited by Surface Interactions
with Polymers}
\date{\today}

\author{\\
{\bf Horacio G. Rotstein}
\thanks{ Volen Center for Complex Systems, 
Brandeis University,
Waltham, MA, 02454-9110, USA.}
{\bf Amy Novick-Cohen}
\thanks{ Department of Mathematics, 
 Technion - IIT, 
Haifa, 32000, Israel.}
{\bf Rina Tannenbaum} 
\thanks{School of Material Sciences and Engineering,
Georgia Institute of Technology, 
Atlanta, GA, 30332-0245, USA.}
}

\def\begeq{\begin{equation}}
\def\endeq{\end{equation}}
\def\begdis{\begin{displaymath}}
\def\enddis{\end{displaymath}}

\def\nd{\noindent}

\begin{document}

\nocite{kn:mas1}
\nocite{kn:leeapa1}
\nocite{kn:apalee1}
\nocite{kn:tomsch1}
\nocite{kn:chered1}
\nocite{kn:gatguc1}
\nocite{kn:andbon1}
\nocite{kn:kalcox1}
\nocite{kn:zakbri1}
\nocite{kn:alfwei1}
\nocite{kn:elkwei1}
\nocite{kn:hamtay1}
\nocite{kn:parnie1}
\nocite{kn:emeehr1}
\nocite{kn:ememah1}
\nocite{kn:grioho1}
\nocite{kn:smiwyc1}
\nocite{kn:tanfle1}
\nocite{kn:alihar1}
\nocite{kn:kerkla1}
\nocite{kn:bru1}
\nocite{kn:reigol1}
\nocite{kn:tangol1}
\nocite{kn:klatan1}
\nocite{kn:imikla1}
\nocite{kn:cotwil1}
\nocite{kn:zif1}
\nocite{kn:smol1}
\nocite{kn:ley1}
\nocite{kn:mac1}
\nocite{kn:zif5}
\nocite{kn:ley2}
\nocite{kn:smol2}
\nocite{kn:rotnov1}
\nocite{kn:bircur1}
\nocite{kn:deg1}
\nocite{kn:jar1}
\nocite{kn:ril1}

\maketitle
\begin{abstract}
The physical and chemical properties of metal nanoparticles differ 
significantly from those of free metal atoms as well as from the properties of
bulk metals, and therefore, they may be viewed as a transition regime between 
the two physical states. Within this nanosize regime, there is a wide 
fluctuation of properties, particularly chemical reactivity, as a function of 
the size, geometry, and electronic state of the metal nanoparticles. In recent
years, great advancements have been made in the attempts to control and 
manipulate the growth of metal particles to pre-specified dimensions. One of 
the main synthetic methods utilized in this endeavor, is the capping of the 
growing clusters with a variety of molecules, e.g. polymers. In this paper we 
attempt to model such a process and show the relationship between the 
concentration of the polymer present in the system and the final metal 
particle size obtained. The theoretical behavior which we obtained is compared
with experimental results for the cobalt-polystyrene system.
\end{abstract}
	The physical and chemical properties of small clusters differ 
significantly from those of free atoms or molecules as well as from
the properties of the bulk surfaces \cite{kn:mas1}-\cite{kn:chered1}, and
hence clusters may be viewed as a transition regime between the two
physical states. Within the cluster regime, there is a wide
fluctuation of properties, particularly chemical reactivity, depending
on the size, geometry, electronic state and packing of the clusters 
\cite{kn:gatguc1}-\cite{kn:parnie1}.

	The motivating physical system for this work is the chemical
synthesis of nanoscale cobalt clusters, via the thermal decomposition
of cobalt carbonyl complexes in solutions containing either pure
solvent  or, in addition, polymers, which will result in a more
viscous reaction environment. The chemical reactivity of metal
clusters is largely determined by two main factors: (a) The size of
the particles; and (b) The oxidation state of the metal clusters. The
ability to control particle size would imply an ability to manipulate
the reactivity and interfacial behavior of the clusters in our
systems. The main factors governing the nucleation and growth of the
metal particles during the decomposition process, are the metal
cluster size; i.e., the number of reactive sites available for 
chemical reactions and the mobility of
the reactive metal clusters and their ability to diffuse through the 
solution and collide with each other. Clearly, as particle size grows,
the mobility of the particles decreases, and a preferred average
pseudo-equilibrium size distribution is reached.
If the medium in which the thermal decompositions take
place is more viscous than a pure solvent, then it is expected that
the upper bound on particle mobility will be reached faster, and hence
the final average particle size will be smaller. 

	In this work, we develop a mathematical model which aims at 
elucidating the 
effect of the initial concentration of the polymer present in the reaction 
solution on the final cobalt particle size obtained via the thermal 
decomposition mechanism, and correlate the model to experimental results. 

	The general chemical synthesis of metal clusters has been designed to
accomodate the manipulation of several process variables, and is best 
described in three stages
(a) The preparation of homogeneous solutions of organometallic complexes in 
a carefully selected solvent; (b) The mixing of these solutions with a 
polymer solution resulting in a viscous medium which contains the 
homogeneously dispersed organometallic complexes; (c) The thermal 
decomposition of the organometallic complexes to form uniform dispersions of 
small particle size.  
The particular system studied in this work consists of cobalt carbonyl 
complexes 
thermally decomposed in the presence of polystyrene (PS)
. It is important to note 
that experimentally, the solvent chosen for the decomposition reactions was 
toluene, which is a good solvent for polystyrene, and therefore it promoted 
the solvation of the polymer over a large compositional range.

	The decomposition of metal carbonyls in various media has been
reported in the literature for the preparation of fine metallic
particles. The thermolysis of transition metal carbonyls in solution
under an inert atmospheres is a well known technique for the
preparation of pure metal powders
\cite{kn:emeehr1}-\cite{kn:tanfle1}. Typically, the clusters formed
following such thermal decomposition are \( 50-250 \) \( \AA \) in diameter,
and hence they belong to the mid-size cluster group, whose diameter
lies within the \( 0-500 \) \( \AA \) range
\cite{kn:alihar1}-\cite{kn:bru1}. Some of the characteristic
properties and advantages of these clusters prepared by the above
method are: (a) The growth of the particles, and hence their final
size, is determined mainly by diffusion distances and the driving
forces between the growing clusters; (b) The chemical and physical
control during synthesis of the clusters allows for the maintenance of
a desired oxidation state and surface cleanliness; (c) The particles
may be polycrystalline \cite{kn:reigol1,kn:tangol1} and hence contain
grain boundaries within the cluster. As a consequence, a large number
of the atoms reside at the interface between these grain boundaries;
(d) Another large fraction of the cluster's atoms  are on or close to
the cluster's surface, and hence these atoms do not order themselves
identically to atoms within a "bulk" material; and (e) There are
various possibilities for the in situ  reaction of these clusters with
a wide range of particles or molecules, and moreover they are small
enough to allow for a homogeneous dispersion within a polymer matrix.

	Since during the decomposition reaction of metal carbonyls highly
reactive intermediate species are formed \cite{kn:klatan1,kn:imikla1},
there are two major pathways for reaction available for these species: 
(a) They can aggregate or fragment to form larger or smaller clusters, 
according to the following general mechanisms respectively:

\begin{eqnarray}
	Co_{j}^{(0)} + Co_{k}^{(0)}	\longrightarrow
	Co_{j+k}^{(0)}, 
						\label{eq:mech1}
	\\
	Co_{j+k}^{(0)}	\longrightarrow
	Co_{j}^{(0)} + Co_{k}^{(0)},
	\nonumber
\end{eqnarray}

\nd where \( j,k = 1,2,\ldots\), and           			 
(b) They can interact with the polymer, according to the following
general mechanism:

\begdis
	Co_{j}^{(0)} + {\mbox PS} \longrightarrow
	Co{\mbox -PS},
							\label{eq:mech2}
\enddis

\nd where \( j = 1,2,\ldots\), resulting in metal attachment processes
which may, in some cases, lead to polymer degradation and/or
crosslinking. All the reactions described here are considered at 
\( 90^{o} C \) and in presence of \( N_{2} \).
The overall chemistry of the system will therefore be a
combination of both pathways, and will be determined by the type of
the reactive groups of the polymer in 
solution  and the nature of the metallic species formed.

	 The overall prototype chemical reaction considered in this
work is given by:

\begeq
	n\ \left[ Co_{x}(CO)_{y}\right] + 
	\mbox{PS} \longrightarrow 
	Co_{m}^{(0)} + 
	\mbox{Co-PS}^{*}, 
							\label{eq:mech3}
\endeq

\nd where  \( ^{*} \) stands for complex compounds and
\( x = 1, 2, 4 \) or \( 6 \) and \( y = 4, 8, 12 \) or \( 16 
\), respectively,  \( y/x \leq 4\), and \( m = j+k \)  in
(\ref{eq:mech1}). In the main reaction, the  clusters are formed from
smaller ones. 
The \( Co-PS^{*} \) complexes formed during this process
are the products of a parallel reaction which competes with the main
coagulation-fragmentation process. 
The extent of the secondary reaction depends on
the reactivity of the cobalt precursor and on the interaction
coefficient between the reactive cobalt species and the polystyrene
molecules. Therefore, the rate of the main coagulation reaction will
depend not only on the concentration of the cobalt precursor and the
rate coefficients of coagulation and fragmentation but also on the
concentration of the available reactive sites on the polystyrene in
the system which are capable of interacting with the cobalt precursor.  

	In order to translate the physical system considered here into
mathematical terms, we first rescale the time variable multiplying it
by the coefficient of reaction between cobalt fragments and polyestyrene,
\( \lambda \), a positive constant, and the initial cluster mass \( C
\), a positive constant. Then we define \( c_{j} \) as the
concentration of \( Co \) clusters of size \( j \) divided by \( C \)
and similarly, \( p \)  as
the concentration of the reactive
sites on the polystyrene, whose initial value is assumed to be a linear
function of the polymer concentration, divided by \( C \).
For the model developed in this paper, we
assumed that the only significant irreversible interaction between
cobalt fragments and polystyrene will occur for \( c_{1} \) with 
a coefficient of reaction \( \lambda \), while for other
clusters, \( j \geq 2\),  the interactions with the polystyrene will be
reversible and weaker \cite{kn:cotwil1}. Therefore, each \( c_{1} \)
particle which is
involved in this interaction will be excluded from further
participation in the main coagulation reaction. Moreover, the chemical
bonding between the \( c_{1} \)  particles and the polystyrene reduces the
number of reactive sites on the polymer and hence, limits the extent
and rate
of the cobalt-polystyrene interaction. It is important to stress that
the overall chemistry of the system and the final particle size
distribution of the cobalt clusters is dependent on the relative
importance and contribution of the two competing reactions. The process 
can therefore be modelled by 
the following system of nonlinear differential equations \cite{kn:rotnov1}:

\begin{eqnarray}
	\dot{c}_{j} = \frac{1}{2} \sum_{k=1}^{j-1}\ [\ R_{j-k,k}\
	c_{j-k}\ c_{k}\ - 
	Q_{j-k,k}\ c_{k}\ ] - 
	\nonumber
	\\
	- \sum_{k=1}^{N-j}\
	[\ R_{j,k}\ c_{j}\ c_{k}\ - Q_{j,k}\ c_{j+k}\ ] - 
	\delta_{1,j}\ c_{1}\ p,
 							\label{eq:def1}\\
	\dot{p} = - c_{1}\ p,
	\ \ \ \ \ \ \ \ \ \ \ \ \ \ \ \ \ \ \ \ \ \ \ \ \ \ \ \ \ \ \
	\ \ \ \ \ \ \ \ \ \ \ 
	\nonumber 
\end{eqnarray}

\nd for \( j = 1, 2, \ldots \), together with initial conditions satisfying \(
\sum_{j=1}^{\infty} j\ c_{j}(0) = 1 \) and \( p(0) = p_{0} \). In 
(\ref{eq:def1}) \( p_{0} \) is the initial dimensional concentration
of reactive sites divided by \( C \), \( R_{jk} \) is the dimensional 
coefficient of coagulation divided by \( \lambda \) and \( Q_{jk} \) 
is the dimensional coefficient of fragmentation divided by \(
\lambda\, C \). All are non-negative and dimensionless. 
If we allow the formation of clusters of all sizes
without any restriction, then \( N = \infty \) in (\ref{eq:def1}).
Since, as we pointed out before, a pseudo-equilibrium state is reached
as a consequence of the fragmentation process and the decrease in
the mobility of large clusters, we assume that the concentration of clusters 
whose size is larger than some finite \( N \) is negligible.

	The  process of metal coagulation and cluster formation (independent 
of the existence of polymer interactions) is a complex series of reactions 
which can be grouped into two categories: (1) The growth of clusters either 
by the formation of atom-atom bonds between two metal fragments and between a
small cluster and a metal fragment, or by the development of surface-surface 
interactions between two large clusters. (2) The 
decomposition of large metal clusters into smaller fragments. 
In order to properly model these processes, it is important to construct the
appropriate reaction rate expressions both for the cluster formation step and
the cluster dissociation step (with or without polymer interaction). 

The simplest coagulation expressions found in the literature which are 
pertinent for our model are: \( R_{jk} = j^{\alpha} k^{\alpha} \) 
(predominantly chemical bond) \cite{kn:zif1}-\cite{kn:ley2}, and 
\( R_{jk} = j^{\alpha} + k^{\alpha} \) (predominantly surface
interactions) \cite{kn:zif1,kn:zif5}, where in both cases 
\( 0 \leq \alpha \leq 1 \). 
In this model, we recognize the fact that there is a 
decline in particle mobility through the medium as a function of increased 
particle size. Moreover, the presence of  the polystyrene molecules in the 
reaction solution contributes to the formation of a viscous medium which will
impose an upper bound on the ability of larger clusters to interact via 
surface-surface interactions, and hence these types of interactions will have
a negligible contribution to the overall coagulation process. These 
assumptions are consistent with the fact that in large clusters the fraction 
of atoms on the surface is small compared to the total number of atoms in the
cluster, and therefore only this small fraction is essentially coordinatively
unsaturated. In effect, we can view large clusters of being thermodynamically
stable in real time. This is not the case for small clusters whose surface 
atoms represent a considerable fraction of the total metal mass. In this case,
the model has to promote the growth of particles formed from smaller 
fragments, because this growth will minimize their free energy by reducing 
the number of coordinatively unsaturated surface atoms.

Therefore, the expression that we propose for the coefficients of coagulation
of the cobalt fragments in our model (with or without polymer interactions) 
is 

\begeq
	R_{j,k} = \frac{j+k}{\sigma}\ j^{\alpha} k^{\alpha}\ +
		  \frac{\sigma}{j+k}\ \frac{\mu}{j^{\beta} k^{\beta}},
							\label{eq:def5}
\endeq

\nd for \( j, k = 1, 2, \ldots \),
where  \( 0 \leq \alpha, \beta \leq 1 \), \( \mu > 0  \) 
is a weighting parameter that denotes the fraction of cobalt clusters whose 
coagulation contribution is negligible, and \( \sigma > 0 \) denotes the 
critical size of  the cobalt clusters beyond which the coagulation rates will
drop as a function of size increase.

The dissociation of cobalt clusters into smaller particles has a direct 
impact on the final equilibrium cluster size distribution of the system. We 
assume that large clusters dissociate preferentially into smaller fragments 
of similar size rather than into fragments with a large variation in
size.
Therefore, the expression that we propose to describe the dissociation of 
large cobalt clusters into smaller fragments is 

\begeq
	Q_{j,k} = \eta\ \frac{j^{\gamma} k^{\gamma}}{j^{\gamma} +
                  k^{\gamma}}
\endeq

\nd for \( j, k = 1, 2, \ldots \), where: \( 0 \leq \gamma \leq 1 \), and 
\( \eta \geq 0 \) is a weighting parameter that denotes the fraction of 
cobalt clusters which undergo dissociation. 

\begin{figure}[ph]
\epsfig{file=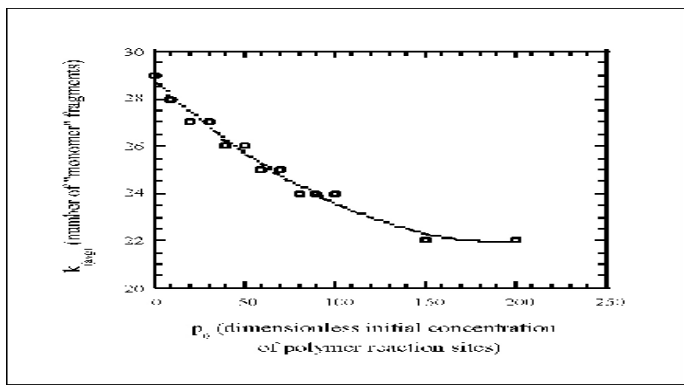,height=5cm,width=9cm,angle=0}    
\caption{The theoretical dependence of cobalt cluster size on the initial 
concentration of reactive sites of polystyrene in the reaction solution, 
\( p_{0} \). Note that the number of reactive sites is directly proportional 
to the polymer concentration for a particular polymer molecular weight.}
\label{theoretical}
\end{figure}

	In order to qualitatively test our model, system
(\ref{eq:def1}) has been solved numerically using a Runge-Kutta method
of order four with 0.005 as time-step and for \( N = 500 \),  
\( \alpha = 0.1 \), \( \beta = 0.1 \), \( \gamma = 1 \), \( \eta =
0.0001 \), \( \sigma = 250 \) and \( \mu = 1.0 \). These values of the 
parameters have been chosen so as the solutions reflect a
pseudo-equilibrium situation in which there is a preferred average
size or, in some cases, preferred average sizes. In Figure \ref{theoretical}
we can see a graph of these preferred average sizes as a function of \( p_{o}
\). 

\begin{figure}[ph]
\epsfig{file=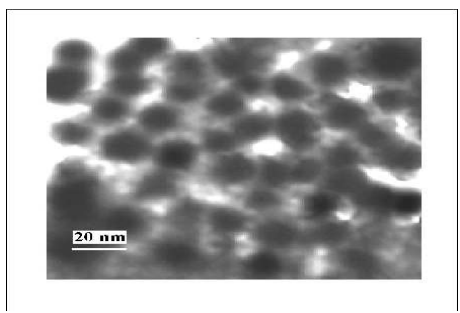,height=4cm,width=6cm,angle=0}  
\caption{The TEM micrograph of zero-valent cobalt particles obtained by the 
thermal decomposition of cobalt carbonyl complexes at \( 90^{o} C\) 
under an inert 
atmosphere in a toluene solution (\(D_{avg}\) = 179.5 \AA).}
\label{cluster-137}
\end{figure}

\begin{figure}[ph]
\epsfig{file=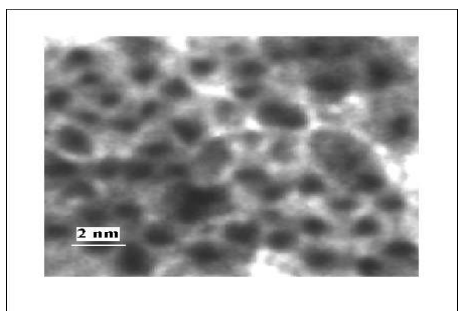,height=4cm,width=6cm,angle=0}    
\caption{The TEM micrograph of zero-valent cobalt particles obtained by the 
thermal decomposition of cobalt carbonyl complexes at \( 90^{o} C \)
under an inert 
atmosphere in  a polystyrene matrix (\(D_{avg}\) = 13.6 \AA).}
\label{cluster-12}
\end{figure}

	Experimental results support the qualitative relationship between
initial concentration of polymer in the reaction solution and the final
particle average size. Experiments were performed by conducting the cobalt
carbonyl decompositions in solutions with different polystyrene
concentrations. A TEM microgrograph picture of cobalt particles obtained
from a decomposition reaction in the absence of PS (solvent only) is shown
in Figure \ref{cluster-137}. The average particle size obtained in this case 
is \(179.5 \AA \).
Another TEM microgrograph picture of cobalt particles obtained from a
decomposition reaction in a PS matrix (~ 85 wt.\% PS in solution) is shown in
Figure \ref{cluster-12}. The average particle size obtained in this case is 
\( 13.8 \AA \).
Clearly, the presence of PS in the reaction medium at these extreme
concentrations has a drastic effect on final average particle size, which
decreases by more that an order of magnitude when the medium consists mainly
of PS. 

\begin{figure}[ph]
\epsfig{file=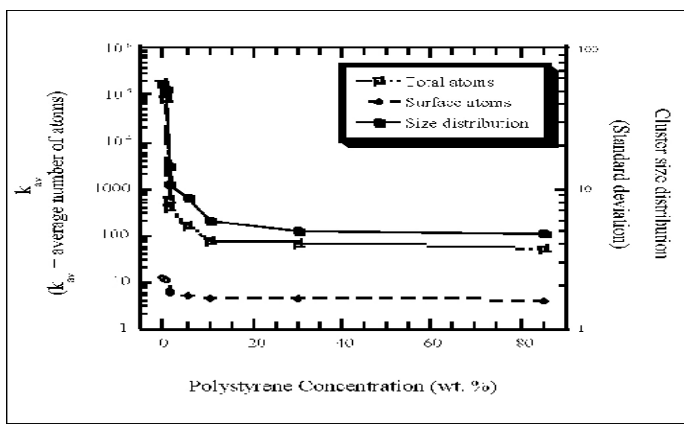,height=5.5cm,width=9cm,angle=0}    
\caption{The experimental realtionship between average cobalt cluster size and
the initial concentration of polystyrene in the reaction solution. The 85 
wt. \% PS
concentration is considered a polystyrene matrix.}
\label{experimental}
\end{figure}

	Figure \ref{experimental} summarizes the experimental data obtained 
for media containing a variety of concentrations of PS. Not only does the 
average final particle size decrease with increasing initial concentration of 
PS, but the average cobalt size distribution decreases as well. This indicates
that the presence of the polymer in the reaction medium prevents the
formation of larger clusters thus inhibiting the coagulation process.


\bibliographystyle{unsrt}
\bibliography{cluster}

\end{document}